\documentclass[showpacs,preprintnumbers,amsmath,amssymb]{revtex4}

\usepackage{graphicx}
\usepackage{dcolumn}
\usepackage[varg]{txfonts}
\usepackage{bm}

\usepackage{color}

\begin{document}

\title{Recursive solutions of Lagrangian perturbation theory}

\author{Takahiko Matsubara}
\email{taka@kmi.nagoya-u.ac.jp}
\affiliation{%
  Department of Physics, Nagoya University, Chikusa, Nagoya 464-8602,
  Japan}%
\affiliation{%
  Kobayashi-Maskawa Institute for the Origin of Particles and the
  Universe, Nagoya University, Chikusa, Nagoya 464-8602, Japan}%

\date{\today}

\begin{abstract}
  In the standard perturbation theory (SPT) of self-gravitating
  Newtonian fluid in an expanding universe, recurrence relations for
  higher-order solutions are well known and play an important role
  both in practical applications and in theoretical investigations.
  The recurrence relations in Lagrangian perturbation theory (LPT),
  however, have not been known for a long time. Recently, two
  different kinds of recurrence relations in LPT have been proposed in
  limited cases. In this paper, we generalize those methods, and most
  generally derive the recurrence relations, which are capable of
  including any initial condition in general models of cosmology. The
  fastest-growing modes in the general relations are identified, and
  simplified recurrence relations with accurate approximation for the
  time dependence are obtained.
\end{abstract}

\pacs{
98.80.-k,
98.65.-r,
04.25.Nx
}
\maketitle


\section{Introduction}

The large-scale structure (LSS) of the Universe is an important source
of information about our Universe. Theoretically understanding
physical origins and statistical properties of LSS is quite important
in cosmology. The driving force for the evolution of LSS is
gravitational instability. Initial density fluctuations are amplified
by the attractive force of gravity and cause the origin of the present
structure. The evolution of LSS is complicated for its nonlinear
nature of dynamics.

The nonlinear structure formation is complicated, and the nonlinear
dynamics in general is quite hard to analytically understand. When the
nonlinearity is considered weak, one can apply the perturbation
theory. The linear theory of gravitational instability describes the
evolution at sufficiently early stages or on sufficiently large scales
of LSS. The linear theory has been quite successful in cosmology. Most
observational signals from cosmic microwave background (CMB) radiation
are understood by the linear theory, because the CMB radiation is
emitted at a sufficiently early time, when the amplitude of density
fluctuations are extremely small. While the amplitudes of density
fluctuations in the present Universe are large on small scales, they
are still small on large scales.

The linear theory of density fluctuations is relatively simple to
analyze, as each Fourier mode independently evolves with time.
However, complicated couplings of mode take place when the
fluctuations becomes nonlinear. Early stages of the mode coupling can
be analytically described by higher-order perturbation theory beyond
the linear approximation \cite{Pee80}. As recent observations of LSS
are large enough, precise descriptions of a weakly nonlinear regime
are important for cosmological analyses. In this respect, the
higher-order perturbation theory attracts much attention these days.

A straightforward way of describing nonlinear perturbations is the
standard perturbation theory (SPT), in which all the perturbation
variables are expanded in Eulerian space \cite{BCGS02}. Along with
recent interests in cosmological perturbation theory, theoretically
various ways of improving the SPT have been proposed, such as the
renormalized perturbation theory (RPT) \cite{CS06a,CS06b}, effective
field theory of large-scale structure (EFTofLSS) \cite{BNSZ12,CHS12},
and many others \cite{Ber13}. The SPT and its extensions fall into a
category of Eulerian perturbation theory. An alternative to the
Eulerian perturbation theory is provided by the Lagrangian
perturbation theory (LPT), in which all the perturbation variables are
expanded in Lagrangian space
\cite{Buc89,Mou91,Buc92,BE93,HBCJ95,Cat95,RW12,Tat13}. The first-order
LPT corresponds to the classic Zel'dovich approximation \cite{Zel70},
which is a generalization of an exact solution in one-dimensional
space \cite{Nov79}. Although the Lagrangian variables are not directly
observable, there exists a systematic way of predicting observable
quantities from LPT \cite{Mat08a,Mat08b,Mat11,RB12,CRW13,PSZ14,Mat14}.

As expected, there are merits and demerits in the SPT and LPT (see,
e.g., Refs.~\cite{CWN09,VSB15}), and they are complementary to each
other. While LPT cannot be extrapolated into a strongly nonlinear
regime beyond the shell-crossing phase, there are several advantages
over the SPT in a weakly nonlinear regime. In cosmological N-body
simulations, it is convenient and customary to use the LPT to set up
the initial conditions. Including the effects of redshift-space
distortions into the higher-order perturbation theory is
straightforward and natural in LPT \cite{TH96,Mat08a}. Most of the
physical models of bias, such as the halo model \cite{MW96}, peaks
model \cite{BBKS}, excursion set peaks \cite{Des13}, etc.~are
Lagrangian bias, in which the bias relations are defined in Lagrangian
space. It is natural to use LPT to describe the evolution of biased
objects in those models. In the formalism of integrated perturbation
theory (iPT) \cite{Mat08b,Mat11,Mat12,Mat14}, the redshift-space
distortions and Lagrangian bias naturally fit into LPT.

One of the striking advantages of SPT over LPT is that there are
well-known recursive solutions of SPT for order-by-order expressions
\cite{GGRJ86}. Those solutions are exact in the Einstein--de~Sitter
(EdS) model of cosmology, and approximate in most of the realistic
models. The recursive solutions are practically crucial to give
predictions of higher-order perturbation theory by numerical
integrations. Evaluating two- or higher-loop corrections to the power
spectrum fully utilizes the recursive solutions of SPT
\cite{CS06a,OTM11,TBNC12,BGK14}.

Until recently, the recursive solutions of LPT had been considered
difficult to obtain, because the displacement fields in LPT are not
irrotational in general, even though the Eulerian velocity fields are
irrotational in the fastest-growing mode \cite{BCGS02}. However,
progress has made by a pioneering work \cite{Ram12} that describes how
recursive solutions for the transverse part of displacement field are
shown to be derived from the irrotational condition for the Eulerian
velocity field. In the same reference, a hybrid procedure is proposed
to derive the recursive solutions for the longitudinal part from the
known recursive solutions of SPT, by using the SPT/LPT correspondence
of perturbation kernels \cite{Mat11,RB12}.

Most recently, in a somehow different context, recurrence relations of
LPT are derived without resorting to SPT \cite{ZF14,RVF15}. Partly
because the main target of the last formalism is to obtain exact
solutions to the nonlinear fluid equations, the expansion parameter is
a time parameter such as the scale factor, instead of the field values
as in the usual LPT. This expansion scheme agrees with that of usual
LPT for the fastest-growing mode in the EdS universe, but not for
other modes and cosmological models. While the new expansion scheme of
LPT based on a time parameter should be useful for solving particular
problems, it is fair to say that recurrence relations in the usual
expansion scheme of LPT are still lacking for general models of
cosmology.

The purpose of this paper is to generalize the above previous work and
seek the recurrence relations within the usual framework of LPT. We
show that they actually exist for any initial condition in general
cosmology, and recursive solutions are explicitly derived. Besides,
irrotational flows in Eulerian space do not need to be assumed to
begin with, and it is explicitly shown that the fastest-growing modes
of LPT automatically result in irrotational flows in Eulerian space.

This paper is organized as follows. In Sec.~\ref{sec:LPT}, the
framework of LPT is reviewed, and fundamental equations are derived
and summarized in convenient forms for our purpose. In
Sec.~\ref{sec:RecRel}, the most general forms of recurrence relations
of LPT are derived both in configuration space and in Fourier space.
In Sec.~\ref{sec:RecApprox}, recurrence relations with a simple and
accurate approximation of time dependence are presented. Conclusions
are given in Sec.~\ref{sec:Concl}. In Appendix~\ref{app:AltForm}, an
alternative form of recurrence relations for the longitudinal part is
presented. In Appendix~\ref{app:Explicit}, recursive solutions up to
seventh order in perturbations are explicitly given.

\section{The Lagrangian perturbation theory}
\label{sec:LPT}

For a pressureless Newtonian self-gravitating fluid, the equation of
motion of the Eulerian comoving coordinates $\bm{x}(t)$ of a fluid
element is given by \cite{Pee80}
\begin{equation}
  \ddot{\bm{x}} + 2H\dot{\bm{x}} = -\frac{1}{a^2}\bm{\nabla}_x\phi(\bm{x},t),
  \label{eq:1-1}
\end{equation}
where a dot represents a Lagrangian derivative of time, $H=\dot{a}/a$
is the time-dependent Hubble parameter, $\bm{\nabla}_x
= \partial/\partial\bm{x}$ is a spatial derivative in Eulerian space,
and $\phi(\bm{x},t)$ is the gravitational potential. The gravitational
potential is related to the density contrast $\delta(\bm{x},t) =
\rho(\bm{x},t)/\bar{\rho}-1$ through a Poisson equation,
\begin{equation}
  {\mbox{\large{$\triangle$}}}_x \phi(\bm{x},t)
  = 4\pi G \bar{\rho} a^2 \delta(\bm{x},t),
  \label{eq:1-2}
\end{equation}
where $\mbox{\large{$\triangle$}}_x = \bm{\nabla}_x^2$ is the
Laplacian operator.

Independent field variables in LPT are the displacement field
$\bm{\varPsi}(\bm{q},t)$, which corresponds to the position difference
of a fluid element between Lagrangian coordinates $\bm{q}$ and
Eulerian coordinates $\bm{x}$ at a given time $t$, i.e.,
\begin{equation}
  \bm{x}(\bm{q},t) = \bm{q} + \bm{\varPsi}(\bm{q},t).
  \label{eq:1-3}
\end{equation}
As mass elements are distributed homogeneously in Lagrangian
coordinates, the Eulerian density field $\rho(\bm{x},t)$ is given by a
continuity relation $\rho(\bm{x},t)d^3x = \bar{\rho}d^3q$, or
$\rho(\bm{x},t) = \bar{\rho}/J(\bm{q},t)$, where
\begin{equation}
  J(\bm{q},t) = \det(\partial\bm{x}/\partial\bm{q})
  \label{eq:1-4}
\end{equation}
is a Jacobian. Taking the divergence and rotation with respect to
Eulerian coordinates, Eq.~(\ref{eq:1-1}) is equivalent to a set of
equations,
\begin{align}
  \bm{\nabla}_x\cdot
  \left( \ddot{\bm{x}} + 2H\dot{\bm{x}} \right)
  &= 4\pi G\bar{\rho} \left(1-\frac{1}{J}\right),
  \label{eq:1-5a}\\
  \bm{\nabla}_x\times
  \left( \ddot{\bm{x}} + 2H\dot{\bm{x}} \right)
  &= 0.
  \label{eq:1-5b}
\end{align}

We define Jacobian matrix elements,
\begin{equation}
  J_{ij}(\bm{q},t) = \frac{\partial x_i}{\partial q_j}
  = \delta_{ij} + \varPsi_{i,j}(\bm{q},t),
  \label{eq:1-6}
\end{equation}
where the comma denotes spatial derivatives with respect to Lagrangian
coordinates, i.e., $\varPsi_{i,j} = \partial \varPsi_i/\partial q_j$.
Eulerian spatial derivatives are given by
\begin{equation}
  \frac{\partial}{\partial x_j} =
  (J^{-1})_{ij} \frac{\partial}{\partial q_i},
  \label{eq:1-7}
\end{equation}
in terms of Lagrangian spatial derivatives, where $(J^{-1})_{ij}$
represents matrix elements of the inverse Jacobian matrix. In terms of
the Levi-Civit\`a symbol $\varepsilon_{ijk}$, the determinant and
inverse matrix are generally given by
\begin{align}
  J &= \frac{1}{6} \varepsilon_{ijk} \varepsilon_{pqr} J_{ip} J_{jq} J_{kr},
  \label{eq:1-8a}\\
  (J^{-1})_{ij}
  &= \frac{1}{2J} \varepsilon_{jkp} \varepsilon_{iqr} J_{kq} J_{pr}.
  \label{eq:1-8b}
\end{align}
The second equation is derived from the first equation and a standard
formula $(J^{-1})_{ij} = J^\textrm{c}_{ji}/J$, where
$J^\textrm{c}_{ij} = \partial J/\partial J_{ij}$ are cofactors of the
Jacobian. The above formulas are quite useful in LPT \cite{RB12}.

In the following, we frequently use a differential operator
\begin{equation}
  \hat{\cal T} \equiv \frac{\partial^2}{\partial t^2}
  + 2H \frac{\partial}{\partial t},
  \label{eq:1-9}
\end{equation}
where the partial derivative is taken with fixed Lagrangian
coordinates, $\partial/\partial t = \partial/\partial t|_{\bm{q}}$.
One should notice that this is {\em not} a first-order differential
operator. For any functions of time, $A(t)$, $B(t)$, $C(t)$, the
product rules of the above operator are given by
\begin{align}
  \hat{\cal T}(AB) &= 
  \hat{\cal T}(A)\,B +  A\,\hat{\cal T}(B)
  + 2 \dot{A} \dot{B},
  \label{eq:1-9-1a}\\
  \hat{\cal T}(ABC) &= 
  \hat{\cal T}(A)\,BC + A\,\hat{\cal T}(B)\,C + AB\,\hat{\cal T}(C)
  + 2 A\dot{B}\dot{C} + 2 \dot{A}B\dot{C} + 2 \dot{A}\dot{B}C,
  \label{eq:1-9-1b}
\end{align}
and so forth. Substituting Eqs.~(\ref{eq:1-3}), (\ref{eq:1-7}) into
Eqs.~(\ref{eq:1-5a}), (\ref{eq:1-5b}), and using the above formulas,
we have
\begin{align}
  \varepsilon_{ijk} \varepsilon_{pqr} J_{ip} J_{jq}
  \left( \hat{\cal T} - \frac{4\pi G\bar{\rho}}{3} \right) J_{kr}
  + 8\pi G \bar{\rho}
  &= 0,
  \label{eq:1-10a}\\
  J_{ij} \varepsilon_{jkp} J_{qk}
  \hat{\cal T} J_{qp}
  &= 0,
  \label{eq:1-10b}
\end{align}
where a contraction identity
$\varepsilon_{ijk} \varepsilon_{ipq} = \delta_{jp} \delta_{kq} -
\delta_{jq} \delta_{kp}$
is used to derive the second equation. With Eq.~(\ref{eq:1-6}), the
above set of equations fully describes the dynamical evolution of the
displacement field $\bm{\varPsi}$. Essentially equivalent equations
are given in Refs.~\cite{Cat95,RB12}, although they assume the
irrotational condition for the velocity field,
$\bm{\nabla}_x\times \bm{v}=\bm{0}$, which is {\em not} assumed here.
It can be shown that the irrotational condition is compatible to the
dynamical equation, Eq.~(\ref{eq:1-10b}), and it is a consequence of
the fastest-growing mode of LPT, as we will explicitly see in the
following sections.

The evolution equations above are nonlinear and it is hopeless to
analytically find a general solution. When the scales of interest
$\lambda$ are sufficiently larger than the typical scales of the
displacement, $|\bm{\varPsi}| \ll \lambda$, it is useful to solve the
nonlinear equations of motion by applying the perturbation theory,
assuming the absolute values of displacement field are small enough.
We expand the displacement field by a perturbation series,
\begin{equation}
  \bm{\varPsi} = \sum_{n=1}^\infty \bm{\varPsi}^{(n)}
  = \bm{\varPsi}^{(1)} + \bm{\varPsi}^{(2)} + \bm{\varPsi}^{(3)} + \cdots,
  \label{eq:1-11}
\end{equation}
where $\bm{\varPsi}^{(n)}$ has the order of $(\bm{\varPsi}^{(1)})^n$.

\section{Recurrence relations}
\label{sec:RecRel}

According to an identity $\bm{\nabla}\times(\bm{\nabla}\times \bm{A})
= \bm{\nabla}(\bm{\nabla}\cdot\bm{A}) - \mbox{\large{$\triangle$}}
\bm{A}$ in the standard vector calculus, the displacement field is
represented in a form,
\begin{equation}
  \bm{\varPsi} =
  \mbox{\large{$\triangle$}}^{-1}
  \left[
    \bm{\nabla}(\bm{\nabla}\cdot\bm{\varPsi})
    - \bm{\nabla}\times(\bm{\nabla}\times \bm{\varPsi})
  \right],
  \label{eq:2-1}
\end{equation}
where $\bm{\nabla} = \partial/\partial\bm{q}$ is the spatial
derivative in Lagrangian coordinates, and
$\mbox{\large{$\triangle$}}^{-1}$ is the inverse operator of the
Laplacian $\mbox{\large{$\triangle$}} = \bm{\nabla}\cdot\bm{\nabla}$.
Specifically, the inverse Laplacian in configuration space, operating
on a given function $F(\bm{q})$, is represented by
\begin{equation}
  \mbox{\large{$\triangle$}}^{-1} F(\bm{q})
  = - \frac{1}{4\pi}
  \int d^3q' \frac{F(\bm{q}')}{|\bm{q}-\bm{q}'|}.
  \label{eq:2-1-1}
\end{equation}
From two kinds of spatial derivatives $\bm{\nabla}\cdot\bm{\varPsi}$
and $\bm{\nabla}\times\bm{\varPsi}$, one can reconstruct the
displacement field by Eq.~(\ref{eq:2-1}). Solutions of the Laplace's
equation, $\mbox{\large{$\triangle$}}\bm{\varPsi}=0$, should not be
added to Eq.~(\ref{eq:2-1}), as we impose the statistical zero mean to
the displacement field, $\langle\bm{\varPsi}\rangle = 0$. We call the
first term in the right-hand side of Eq.~(\ref{eq:2-1}) the
longitudinal part, and the second term the transverse part. In
Lagrangian dynamics, these two parts are generally coupled to each
other.

\subsection{Longitudinal part}

Substituting Eq.~(\ref{eq:1-6}) into Eqs.~(\ref{eq:1-10a}), we have
\begin{equation}
  \left( \hat{\cal T} - 4\pi G\bar{\rho} \right)
  \varPsi_{i,i}
  = - \varepsilon_{ijk} \varepsilon_{ipq} \varPsi_{j,p}
  \left( \hat{\cal T} - 2\pi G\bar{\rho} \right) \varPsi_{k,q}
  - \frac{1}{2} \varepsilon_{ijk} \varepsilon_{pqr}
  \varPsi_{i,p} \varPsi_{j,q}
  \left( \hat{\cal T} - \frac{4\pi G}{3}\bar{\rho} \right) \varPsi_{k,r}.
  \label{eq:2-2}
\end{equation}
While the left-hand side linearly depends on the displacement field,
the right-hand side consists of higher-order terms. The structure of
this equation is the basis of the recurrence relations. Regarding the
right-hand side as a source function, Eq.~(\ref{eq:2-2}) has a form of
inhomogeneous linear differential equation,
\begin{equation}
  \left(\hat{\cal T} - 4\pi G\bar{\rho}\right) g(t) = F(t).
  \label{eq:2-2-1}
\end{equation}

The general solution of Eq.~(\ref{eq:2-2-1}) is found by a standard
method of ordinary differential equations as follows. The homogeneous
equation of Eq.~(\ref{eq:2-2-1}) is exactly the same as the linearized
evolution equation of the Eulerian density contrast, $\ddot{\delta} +
2H\dot{\delta} - 4\pi G \bar{\rho} \delta = 0$, and thus the two
independent solutions of the homogeneous equation are the growing mode
solution $D_+(t)$ and decaying mode solution $D_-(t)$ \cite{Pee80}.
For example, $D_+(t) \propto t^{2/3}$, $D_-(t) \propto t^{-1}$ in the
EdS model, and some analytic solutions are known for several
cosmological models. The growing mode solution $D_+(t)$ is also known
as the linear growth factor. The general solution of the inhomogeneous
differential equation, Eq.~(\ref{eq:2-2-1}), is given by
\begin{equation}
  g(t) =
  C_1 D_+(t) + C_2 D_-(t)
  + \int_{t_\textrm{in}}^t G(t,t') F(t') dt',
  \label{eq:2-2-4}
\end{equation}
where $C_1$ and $C_2$ are integration constants, $t_\textrm{in}$ is
the initial time, and
\begin{equation}
  G(t,t') \equiv 
  \frac{D_+(t)D_-(t') - D_-(t)D_+(t')}
  {\dot{D}_+(t')D_-(t') - \dot{D}_-(t')D_+(t')}.
  \label{eq:2-2-5}
\end{equation}
The last term of Eq.~(\ref{eq:2-2-4}) is a particular solution of the
inhomogeneous equation. We denote this particular solution as an
inverse of the differential operator,
\begin{equation}
  \left(\hat{\cal T} - 4\pi G\bar{\rho}\right)^{-1} F(t)
  \equiv \int_{t_\textrm{in}}^t G(t,t') F(t') dt',
  \label{eq:2-2-6}
\end{equation}
which is a linear operator.

Using the notations above, a formal solution of Eq.~(\ref{eq:2-2}) is
given by
\begin{equation}
  \bm{\nabla}\cdot\bm{\varPsi}
  = D_+(t) A_+ + D_-(t) A_-
  - \left(\hat{\cal T} - 4\pi G\bar{\rho}\right)^{-1}
  \left[
      \varepsilon_{ijk} \varepsilon_{ipq} \varPsi_{j,p}
      \left( \hat{\cal T} - 2\pi G\bar{\rho} \right) \varPsi_{k,q}
      + \frac{1}{2} \varepsilon_{ijk} \varepsilon_{pqr}
      \varPsi_{i,p} \varPsi_{j,q}
      \left( \hat{\cal T} - \frac{4\pi G}{3}\bar{\rho} \right)
      \varPsi_{k,r}
  \right],
  \label{eq:2-2-7}
\end{equation}
where $A_\pm$ depends only on Lagrangian coordinates. The first two
terms of the right-hand side give the linear solution of the
longitudinal part,
\begin{equation}
  \bm{\nabla}\cdot\bm{\varPsi}^{(1)}
  = D_+(t) A_+(\bm{q}) + D_-(t) A_-(\bm{q}).
  \label{eq:2-2-8}
\end{equation}
The functions $A_\pm(\bm{q})$ are determined by initial conditions.
Substituting the perturbation series of Eq.~(\ref{eq:1-11}) into
Eq.~(\ref{eq:2-2-7}) and extracting a component of particular order $n
\geq 2$, we have
\begin{multline}
  \bm{\nabla}\cdot\bm{\varPsi}^{(n)}
  = - \sum_{m_1 + m_2 = n}
  \left( \hat{\cal T} - 4\pi G\bar{\rho} \right)^{-1}
  \left[
      \varepsilon_{ijk} \varepsilon_{ipq} \varPsi^{(m_1)}_{j,p}
      \left( \hat{\cal T} - 2\pi G\bar{\rho} \right) \varPsi^{(m_2)}_{k,q}
  \right]
\\
  - \frac{1}{2}
  \sum_{m_1+m_2+m_3=n}
  \left( \hat{\cal T} - 4\pi G\bar{\rho} \right)^{-1}
  \left[
      \varepsilon_{ijk} \varepsilon_{pqr}
      \varPsi^{(m_1)}_{i,p} \varPsi^{(m_2)}_{j,q}
      \left( \hat{\cal T} - \frac{4\pi G}{3}\bar{\rho} \right)
      \varPsi^{(m_3)}_{k,r}
  \right].
  \label{eq:2-3}
\end{multline}
The right-hand side consists of only lower-order perturbations up to
order $n-1$. When the lower-order solutions
$\bm{\varPsi}^{(1)},\ldots,\bm{\varPsi}^{(n-1)}$ are known, the
longitudinal part of $n$th order displacement field
$\bm{\varPsi}^{(n)}$ is obtained from this equation. The above
Eq.~(\ref{eq:2-3}) alone is not sufficient to determine the
displacement field, because it lacks the transverse part. On the
right-hand side, both longitudinal and transverse parts of lower-order
perturbations enter, and thus both parts of displacement field are
coupled to each other. We need a similar equation for the transverse
part, which will be derived in the next subsection. There is an
alternative but equivalent representation of Eq.~(\ref{eq:2-3}), which
is described in Appendix~\ref{app:AltForm}.

Because products of Levi-Civit\`a symbol can be represented by
products of the Kronecker delta, Eq.~(\ref{eq:2-3}) can be
equivalently expressed without Levi-Civit\`a symbols. Specifically, the
identities
\begin{equation}
  \varepsilon_{ijk}\varepsilon_{ipq} =
  \left|
    \begin{matrix}
      \delta_{jp} & \delta_{jq} \\
      \delta_{kp} & \delta_{kq} \\
    \end{matrix}
  \right|,
  \quad
  \varepsilon_{ijk}\varepsilon_{pqr}
  =
  \left|
    \begin{matrix}
      \delta_{ip} & \delta_{iq} &  \delta_{ir} \\
      \delta_{jp} & \delta_{jq} &  \delta_{jr} \\
      \delta_{kp} & \delta_{kq} &  \delta_{kr}
    \end{matrix}
  \right|,
  \label{eq:2-4}
\end{equation}
imply
\begin{align}
  \varepsilon_{ijk}\varepsilon_{ipq} A_{jp} B_{kq}
  &=
  \textrm{Tr}A\,\textrm{Tr}B - \textrm{Tr}(AB),
  \label{eq:2-5a}\\
  \varepsilon_{ijk}\varepsilon_{pqr} A_{ip} B_{jq} C_{kr}
  &=
  \textrm{Tr}A\,\textrm{Tr}B\,\textrm{Tr}C
  - \textrm{Tr}A\,\textrm{Tr}(BC)
  - \textrm{Tr}B\,\textrm{Tr}(CA)
  - \textrm{Tr}C\,\textrm{Tr}(AB)
  + \textrm{Tr}(ABC) + \textrm{Tr}(ACB).
  \label{eq:2-5b}
\end{align}
Substituting these identities into Eq.~(\ref{eq:2-3}), Levi-Civit\`a
symbols can be eliminated in the expression.

\subsection{Transverse part}

Noting Eq.~(\ref{eq:1-10b}) has the form of $J_{ij} X_j = 0$, and
multiplying the inverse Jacobian matrix from the left, we have
$X_i=0$. Therefore Eq.~(\ref{eq:1-10b}) is equivalent to
\begin{equation}
 \varepsilon_{ijk} J_{pj} \hat{\cal T} J_{pk}
  = 0.
  \label{eq:2-10}
\end{equation}
Substituting Eq.~(\ref{eq:1-6}) into the above equation, we have
\begin{equation}
  \hat{\cal T} \bm{\nabla}\times\bm{\varPsi}
  = \bm{\nabla}\varPsi_i \times \hat{\cal T} \bm{\nabla}\varPsi_i.
  \label{eq:2-11}
\end{equation}
An apparent solution of the homogeneous equation $\hat{\cal T}g(t)=0$
is a constant, and another solution is
\begin{equation}
  E_-(t) \equiv \int_t^\infty \frac{dt}{a^2}
  = \int_{a(t)}^\infty \frac{da}{a^3H},
  \label{eq:2-11-1}
\end{equation}
which is a decaying function of time. In the EdS universe, $E_-(t)
\propto t^{-1/3}$. From the above solutions of the homogeneous
equation, an inverse operator of $\hat{\cal T}$ is constructed as
\begin{equation}
  \hat{\cal T}^{-1} F(t) =
  \int_{t_\textrm{in}}^t 
  a^2(t')\left[E_-(t)-E_-(t')\right] F(t') dt'
  = \int_{t_\textrm{in}}^t 
  a^2(t')
  \left[\int_t^{t'} \frac{dt''}{a^2(t'')}\right] F(t') dt'.
  \label{eq:2-11-2}
\end{equation}

The formal solution of Eq.~(\ref{eq:2-11}) is then given by
\begin{equation}
  \bm{\nabla}\times\bm{\varPsi} =
  \bm{B}_0 + E_-(t)\bm{B}_-
  + \hat{\cal T}^{-1}
  \left(
      \bm{\nabla}\varPsi_i \times
      \hat{\cal T} \bm{\nabla}\varPsi_i
  \right),
  \label{eq:2-11-3}
\end{equation}
where $\bm{B}_{0,-}(\bm{q})$ are integration constants which depend
only on Lagrangian coordinates and are divergence-free,
$\bm{\nabla}\cdot\bm{B}_{0,-}=0$. The first two terms of the
right-hand side give the linear solution of the transverse part,
\begin{equation}
  \bm{\nabla}\times\bm{\varPsi}^{(1)} =
  \bm{B}_0(\bm{q}) + E_-(t)\bm{B}_-(\bm{q}).
  \label{eq:2-11-4}
\end{equation}
The divergence-free vectors $\bm{B}_{0,-}(\bm{q})$ are determined by
initial conditions. Substituting the perturbation series of
Eq.~(\ref{eq:1-11}) and extracting a component of particular order $n
\geq 2$, we have
\begin{equation}
  \bm{\nabla}\times\bm{\varPsi}^{(n)} =
  \sum_{m_1+m_2=n}
  \hat{\cal T}^{-1}
  \left(
      \bm{\nabla} \varPsi^{(m_1)}_i
      \times \hat{\cal T} \bm{\nabla} \varPsi^{(m_2)}_i
  \right).
  \label{eq:2-12}
\end{equation}
The right-hand side consists of only lower-order perturbations up to
order $n-1$, and thus $n$th-order perturbations of the transverse part
are given only by lower-order perturbations. As in the case of
Eq.~(\ref{eq:2-3}), both the longitudinal and transverse parts of
lower-order perturbations enter in the right-hand side.

Eqs.~(\ref{eq:2-3}) and (\ref{eq:2-12}), together with
Eq.~(\ref{eq:2-1}) of each order,
\begin{equation}
  \bm{\varPsi}^{(n)} =
  \mbox{\large{$\triangle$}}^{-1}
  \left[
    \bm{\nabla}\left(\bm{\nabla}\cdot\bm{\varPsi}^{(n)}\right)
    - \bm{\nabla}\times\left(\bm{\nabla}\times \bm{\varPsi}^{(n)}\right)
  \right],
  \label{eq:2-13}
\end{equation}
are the closed set of general recurrence relations from which
recursive solutions of LPT can be derived. 

\subsection{Seed values: First-order solution and initial condition}

The seed values for the recursive solutions are given by the
first-order solutions, Eqs.~(\ref{eq:2-2-8}), (\ref{eq:2-11-4}). From
those equations and Eq.~(\ref{eq:2-13}) with $n=1$, the general
solution of first order is given by
\begin{equation}
  \bm{\varPsi}^{(1)}(\bm{q},t) =
  \mbox{\large{$\triangle$}}^{-1}
  \left\{
    \bm{\nabla}\left[D_+(t)A_+(\bm{q}) + D_-(t)A_-(\bm{q})\right]
    + \bm{\nabla}\times
    \left[\bm{B}_0(\bm{q}) + E_-(t)\bm{B}_-(\bm{q})\right]
  \right\}.
  \label{eq:2-17}
\end{equation}
As the functions $\bm{B}_{0,-}(\bm{q})$ are divergence free, the set
of functions $A_\pm(\bm{q})$, $\bm{B}_{0,-}(\bm{q})$ has 6 degrees of
freedom at each Lagrangian point. They are completely determined by
the initial condition of displacement field, $\bm{\varPsi}_\textrm{in}
= \bm{\varPsi}^{(1)}(\bm{q},t_\textrm{in})$ and
$\dot{\bm{\varPsi}}_\textrm{in} =
\dot{\bm{\varPsi}}^{(1)}(\bm{q},t_\textrm{in})$. However, physical
degrees of freedom in an initial condition for a fluid element are
just 4, instead of 6. They are the initial values of density contrast
$\delta_\textrm{in}(\bm{q}) \equiv \delta(\bm{q},t_\textrm{in})$ and
peculiar velocity $\bm{v}_\textrm{in}(\bm{q}) \equiv
\bm{v}(\bm{q},t_\textrm{in})$ at an initial time $t_\textrm{in}$. Thus
the initial conditions for the displacement field have physically
redundant degrees of freedom.

To resolve the redundancy, we note that the initial density contrast
and peculiar velocity are given by
\begin{equation}
  \delta_\textrm{in} =
  -\bm{\nabla}\cdot\bm{\varPsi}_\textrm{in}, \qquad
  \bm{v}_\textrm{in} = a_\textrm{in} \dot{\bm{\varPsi}}_\textrm{in},
  \label{eq:2-18}
\end{equation}
where $a_\textrm{in} = a(t_\textrm{in})$ is the value of the scale
factor at the initial time, and we assume that the initial time is
sufficiently early and the density contrast is well within the linear
regime. It is obvious that only the functions $A_\pm(\bm{q})$ and
$\bm{B}_-(\bm{q})$ can be determined by $\delta_\textrm{in}$ and
$\bm{v}_\textrm{in}$, and the function $\bm{B}_0(\bm{q})$ remains
undetermined. Therefore, the initial conditions for the
constant-transverse mode $\bm{B}_0$ cannot be associated with any
physical quantity. This is natural, because a time-invariant rotation
of displacement field is just a relabeling of the Lagrangian
coordinates of fluid elements, without changing the density and
velocity; only the time-varying rotation of displacement field is
physically relevant to the vorticity. Thus the function $\bm{B}_0$ can
be considered as the ``gauge'' degrees of freedom in the initial
condition, and we can freely choose this function without affecting
any physical quantity (essentially the same argument is found in
Ref.~\cite{FS78} in nonexpanding background space). This function
$\bm{B}_0$ has 2 degrees of freedom because of the divergence-free
condition, and accounts for the redundancy of the initial conditions
described above. The simplest and most natural choice is obviously
$\bm{B}_0 = \bm{0}$. The arbitrariness of this gauge mode does not
affect the fastest-growing mode of displacement field.

The expansion of the Universe at the initial time is well described by
the EdS universe, so we have $D_+(t_\textrm{in}) \propto
{t_\textrm{in}}^{2/3}$, $D_-(t_\textrm{in}) \propto
{t_\textrm{in}}^{-1}$, $E_-(t_\textrm{in}) \propto
{t_\textrm{in}}^{-1/3}$, and
$\dot{D}_+(t_\textrm{in})/D_+(t_\textrm{in}) = 2/3t_\textrm{in}$,
$\dot{D}_-(t_\textrm{in})/D_-(t_\textrm{in}) = -1/t_\textrm{in}$,
$\dot{E}_+(t_\textrm{in})/E_+(t_\textrm{in}) = - 1/3t_\textrm{in}$.
Using these relations, Eq.~(\ref{eq:2-18}) completely determines the
functions $A_\pm(\bm{q})$, $\bm{B}_-(\bm{q})$. As a result,
Eq.~(\ref{eq:2-17}) can be represented by
\begin{equation}
  \bm{\varPsi}^{(1)} =
  - \mbox{\large{$\triangle$}}^{-1}
  \left[
    \frac{3}{5} \frac{D_+(t)}{D_+(t_\textrm{in})}
    \bm{\nabla}
    \left(
        \delta_\textrm{in}
        - \frac{2}{3}
        \frac{\bm{\nabla}\cdot\bm{v}_\textrm{in}}{a_\textrm{in}H_\textrm{in}}
    \right)
    + \frac{2}{5} \frac{D_-(t)}{D_-(t_\textrm{in})}
    \bm{\nabla}
    \left(
        \delta_\textrm{in}
        + \frac{\bm{\nabla}\cdot\bm{v}_\textrm{in}}{a_\textrm{in}H_\textrm{in}}
    \right)
    - 2 \frac{E_-(t)}{E_-(t_\textrm{in})}
    \frac{\bm{\nabla}\times(\bm{\nabla}\times\bm{v}_\textrm{in})}
    {a_\textrm{in}H_\textrm{in}}
  \right],
  \label{eq:2-19}
\end{equation}
where $H_\textrm{in} = 2/3t_\textrm{in}$ is the Hubble parameter at
the initial time, and we choose $\bm{B}_0=\bm{0}$ as described above.
The Eulerian counterpart of Eq.~(\ref{eq:2-19}) can be found in \S 15
of Ref.~\cite{Pee80}. The first term with a linear growth factor $D_+$
is the fastest-growing mode. In the most general case,
Eq.~(\ref{eq:2-19}) can be used as seed values for the recurrence
relations, Eqs.~(\ref{eq:2-3}), (\ref{eq:2-12}), and (\ref{eq:2-13}).
Keeping only the fastest-growing mode is sufficiently accurate in most
practical applications. Denoting
\begin{equation}
  \delta_\textrm{L}(\bm{q},t) \equiv
  \frac{3}{5} \frac{D_+(t)}{D_+(t_\textrm{in})}
  \left[
      \delta_\textrm{in}(\bm{q})
      - \frac{2}{3}
      \frac{\bm{\nabla}\cdot\bm{v}_\textrm{in}(\bm{q})}
      {a_\textrm{in}H_\textrm{in}}
  \right],
  \label{eq:2-19-1}
\end{equation}
and keeping only the fastest-growing mode, the Zel'dovich
approximation \cite{Zel70},
$\bm{\varPsi}^{(1)} = - \bm{\nabla} \mbox{\large{$\triangle$}}^{-1}
\delta_\textrm{L}$,
is recovered. Adopting the Zel'dovich approximation as the seed values
for the recurrence relations, Eqs.~(\ref{eq:2-3}), (\ref{eq:2-12}),
(\ref{eq:2-13}) of $n\geq 2$, the resultant recursive solutions are
also the fastest-growing mode of higher-order perturbations.

\subsection{Representations in Fourier space}

Applying the Fourier transform to the displacement field,
\begin{equation}
  \bm{\varPsi}(\bm{q},t) =
  \int \frac{d^3k}{(2\pi)^3}
  e^{i\bm{k}\cdot\bm{q}} \tilde{\bm{\varPsi}}(\bm{k},t),
  \label{eq:2-20}
\end{equation}
Eq.~(\ref{eq:2-3}), (\ref{eq:2-12}) and (\ref{eq:2-13}) are
transformed to
\begin{align}
  \bm{k}\cdot\tilde{\bm{\varPsi}}^{(n)}(\bm{k})
  &= -i \sum_{m_1 + m_2 = n}
  \int_{\bm{k}_{12}=\bm{k}}
  (\bm{k}_1\times\bm{k}_2)
  \cdot
  \left( \hat{\cal T} - 4\pi G\bar{\rho} \right)^{-1}
  \left[
      \tilde{\bm{\varPsi}}^{(m_1)}(\bm{k}_1)\times
      \left( \hat{\cal T} - 2\pi G\bar{\rho} \right)
      \tilde{\bm{\varPsi}}^{(m_2)}(\bm{k}_2)
  \right]
  \nonumber\\
  & \quad
  + \frac{1}{2} \sum_{m_1+m_2+m_3=n}
  \int_{\bm{k}_{123}=\bm{k}}
  \left[
      \bm{k}_1\cdot(\bm{k}_2\times\bm{k}_3)
  \right]
  \left( \hat{\cal T} - 4\pi G\bar{\rho} \right)^{-1}
  \left\{
      \tilde{\bm{\varPsi}}^{(m_1)}(\bm{k}_1) \cdot
      \left[
          \tilde{\bm{\varPsi}}^{(m_2)}(\bm{k}_2)\times
          \left( \hat{\cal T} - \frac{4\pi G}{3}\bar{\rho} \right)
          \tilde{\bm{\varPsi}}^{(m_3)}(\bm{k}_3)
      \right]
  \right\}
  \Biggr]
  \label{eq:2-21a}\\
  \bm{k}\times\tilde{\bm{\varPsi}}^{(n)}(\bm{k})
  &= i \sum_{m_1+m_2=n}
  \int_{\bm{k}_{12}=\bm{k}} (\bm{k}_1\times\bm{k}_2)
  \hat{\cal T}^{-1}
  \left[
    \tilde{\bm{\varPsi}}^{(m_1)}(\bm{k}_1) \cdot
    \hat{\cal T}\tilde{\bm{\varPsi}}^{(m_2)}(\bm{k}_2)
  \right],
  \label{eq:2-21b}\\
  \tilde{\bm{\varPsi}}^{(n)}(\bm{k}) &= 
  \frac{1}{k^2}
  \left\{
    \bm{k}\left[\bm{k}\cdot\tilde{\bm{\varPsi}}^{(n)}(\bm{k})\right]
    - \bm{k}\times[\bm{k}\times\tilde{\bm{\varPsi}}^{(n)}(\bm{k})]
  \right\},
  \label{eq:2-21c}
\end{align}
where we adopt notations,
\begin{equation}
  \bm{k}_{1\cdots n} \equiv \bm{k}_1 + \cdots + \bm{k}_n,
  \qquad
  \int_{\bm{k}_{1\cdots n}=\bm{k}} \cdots \equiv
  \int \frac{d^3k_1}{(2\pi)^3}\cdots\frac{d^3k_n}{(2\pi)^3}
  \delta_\textrm{D}^3(\bm{k}_{1\cdots n}-\bm{k}) \cdots .
  \label{eq:2-23}
\end{equation}
The seed values for the above recurrence relations are given by the
Fourier transform of Eq.~(\ref{eq:2-13}), i.e., 
\begin{equation}
  \tilde{\bm{\varPsi}}^{(1)}(\bm{k}) =
  \frac{i}{k^2}
  \left[
    \frac{3}{5} \frac{D_+(t)}{D_+(t_\textrm{in})}
    \bm{k}
    \left(
        \tilde{\delta}_\textrm{in}
        - \frac{2}{3}
        \frac{i\bm{k}\cdot\tilde{\bm{v}}_\textrm{in}}{a_\textrm{in}H_\textrm{in}}
    \right)
    + \frac{2}{5} \frac{D_-(t)}{D_-(t_\textrm{in})}
    \bm{k}
    \left(
        \tilde{\delta}_\textrm{in}
        + \frac{i\bm{k}\cdot\tilde{\bm{v}}_\textrm{in}}{a_\textrm{in}H_\textrm{in}}
    \right)
    - 2 \frac{E_-(t)}{E_-(t_\textrm{in})}
    \frac{i\bm{k}\times\left(\bm{k}\times\tilde{\bm{v}}_\textrm{in}\right)}
    {a_\textrm{in}H_\textrm{in}}
  \right].
  \label{eq:2-23-1}
\end{equation}
Although the time dependence is dropped from the argument of
$\tilde{\bm{\varPsi}}^{(n)}$ for notational simplicity, it actually
does depend on the time variable.

The above equations are the most general form of recurrence relations
in Fourier space. When only the fastest-growing mode is considered,
the first-order solution is given by
\begin{equation}
  \tilde{\bm{\varPsi}}^{(1)}(\bm{k})
  = \frac{i\bm{k}}{k^2} D(t)\delta_0(\bm{k}),
  \label{eq:2-24-1}
\end{equation}
where $D(t) = D_+(t)/D_+(t_0)$ is the linear growth factor normalized
at the present time as $D(t_0)=1$, and $\delta_0(\bm{k}) =
\tilde{\delta}_\textrm{L}(\bm{k},t_0)$ is the linear density contrast
at the present time. Considering only the fastest-growing mode, it is
possible to represent the displacement field of each order as
\begin{equation}
  \tilde{\bm{\varPsi}}^{(n)}(\bm{k},t) = \frac{i}{N!}
  \int_{\bm{k}_{1\cdots n}=\bm{k}}
  \tilde{\bm{L}}_n(\bm{k}_1,\ldots,\bm{k}_n;t)
  \delta_0(\bm{k}_1)\cdots\delta_0(\bm{k}_n),
  \label{eq:2-24}
\end{equation}
where $\tilde{\bm{L}}_n$ are time-dependent Fourier kernels. For
$n=1$, we have $\tilde{\bm{L}}(\bm{k};t) = D(t)\bm{k}/k^2$.

In the following, we define and use the functions,
\begin{align}
  \tilde{S}_n(\bm{k}_1,\ldots,\bm{k}_n;t) &\equiv
  \bm{k}_{1\cdots n}\cdot
  \tilde{\bm{L}}_n(\bm{k}_1,\ldots,\bm{k}_n;t), \qquad
  \label{eq:2-25a}\\
  \tilde{\bm{T}}_n(\bm{k}_1,\ldots,\bm{k}_n;t) &\equiv
  - \bm{k}_{1\cdots n}\times
  \tilde{\bm{L}}_n(\bm{k}_1,\ldots,\bm{k}_n;t).
  \label{eq:2-25b}
\end{align}
Substituting Eq.~(\ref{eq:2-24}) into
Eqs.~(\ref{eq:2-21a})--(\ref{eq:2-21c}), the recurrence relations for
the fastest-growing mode of Fourier kernels are obtained as
\begin{align}
  \tilde{S}_n(\bm{k}_1,\ldots,\bm{k}_n,t)
   &= \sum_{m_1 + m_2 = n} \frac{n!}{m_1! m_2!}
   (\bm{k}_{1\cdots m_1}\times\bm{k}_{(m_1+1)\cdots n})
  \cdot
  \left( \hat{\cal T} - 4\pi G\bar{\rho} \right)^{-1}
  \left[
    \tilde{\bm{L}}_{m_1}(\bm{k}_1,\ldots,\bm{k}_{m_1};t) \times
    \left( \hat{\cal T} - 2\pi G\bar{\rho} \right)
    \tilde{\bm{L}}_{m_2}(\bm{k}_{m_1+1},\ldots,\bm{k}_n;t)
  \right]
\nonumber\\
  & \quad
   - \frac{1}{2} \sum_{m_1+m_2+m_3=n}
   \frac{n!}{m_1! m_2! m_3!}
  \left[
    \bm{k}_{1\cdots m_1}\cdot(\bm{k}_{m_1\cdots(m_1+m_2)}
    \times\bm{k}_{(m_1+m_2+1)\cdots n})
  \right]
\nonumber\\
  & \qquad
  \left( \hat{\cal T} - 4\pi G\bar{\rho} \right)^{-1}
  \left\{
    \tilde{\bm{L}}_{m_1}(\bm{k}_1,\ldots,\bm{k}_{m_1};t) \cdot
    \left[
      \tilde{\bm{L}}_{m_2}(\bm{k}_{m_1+1},\ldots,\bm{k}_{m_1+m_2};t)\times
      \left( \hat{\cal T} - \frac{4\pi G}{3}\bar{\rho} \right)
      \tilde{\bm{L}}_{m_3}(\bm{k}_{m_1+m_2+1},\ldots,\bm{k}_n;t)
    \right]
  \right\},
  \label{eq:2-26a}\\
  \tilde{\bm{T}}_n(\bm{k}_1,\ldots,\bm{k}_n;t)
  &= \sum_{m_1 + m_2 = n}
  \frac{n!}{m_1! m_2!}
  (\bm{k}_{1\cdots m_1}\times\bm{k}_{(m_1+1)\cdots n})
  \hat{\cal T}^{-1}
  \left[
    \tilde{\bm{L}}_{m_1}(\bm{k}_1,\ldots,\bm{k}_{m_1};t) \cdot
    \hat{\cal T}
    \tilde{\bm{L}}_{m_2}(\bm{k}_{m_1+1},\ldots,\bm{k}_n;t)
  \right],
  \label{eq:2-26b}\\
  \tilde{\bm{L}}_n(\bm{k}_1,\ldots,\bm{k}_n;t)
  &=  \frac{1}{{k_{1\cdots n}}^2} 
  \left[
    \bm{k}_{1\cdots n} \tilde{S}_n(\bm{k}_1,\ldots,\bm{k}_n;t)
    + \bm{k}_{1\cdots n}\times
      \tilde{\bm{T}}_n(\bm{k}_1,\ldots,\bm{k}_n;t)
  \right].
  \label{eq:2-26c}
\end{align}
The right-hand sides of the above equations are not symmetric with
respect to their wave vectors $\bm{k}_1,\ldots,\bm{k}_n$, and so are
the perturbation kernels obtained by the recurrence relations. Those
wave vectors are interchangeable in Eq.~(\ref{eq:2-24}), and thus only
symmetrized kernels are physically relevant. The symmetrized kernels
are obtained from unsymmetric kernels by a symmetrization procedure
\begin{equation}
  \tilde{\bm{L}}^\textrm{sym.}_n(\bm{k}_1,\ldots,\bm{k}_n;t)
  \equiv
  \frac{1}{n!} \sum_{p \in {\cal S}_n}
  \tilde{\bm{L}}_n(\bm{k}_{p(1)},\ldots,\bm{k}_{p(n)};t),
  \label{eq:2-28}
\end{equation}
where the summation is taken for all the possible permutations ${\cal S}_n$
of the arguments. Corresponding to the relations of
Eqs.~(\ref{eq:2-4})--(\ref{eq:2-5b}), there are vector identities
\begin{align}
  (\bm{k}\times\bm{k}')\cdot(\bm{L}\times\bm{L}')
  &=
  \left|
    \begin{matrix}
      \bm{k}\cdot\bm{L} & \bm{k}\cdot\bm{L}' \\
      \bm{k}'\cdot\bm{L} & \bm{k}'\cdot\bm{L}'
    \end{matrix}
  \right|,
  \label{eq:2-29a}\\
  \left[\bm{k}\cdot(\bm{k}'\times\bm{k}'')\right]
  \left[\bm{L}\cdot(\bm{L}'\times\bm{L}'')\right]
  &=
  \left|
    \begin{matrix}
      \bm{k}\cdot\bm{L} & \bm{k}\cdot\bm{L}' & \bm{k}\cdot\bm{L}'' \\
      \bm{k}'\cdot\bm{L} & \bm{k}'\cdot\bm{L}' & \bm{k}'\cdot\bm{L}'' \\
      \bm{k}''\cdot\bm{L} & \bm{k}''\cdot\bm{L}' & \bm{k}''\cdot\bm{L}''
    \end{matrix}
  \right|,
  \label{eq:2-29b}
\end{align}
which provide an alternative expression for Eq.~(\ref{eq:2-26a}).

\section{Recurrence relations with approximate time dependence}
\label{sec:RecApprox}

The recurrence relations derived above, Eqs.~(\ref{eq:2-3}),
(\ref{eq:2-12}), (\ref{eq:2-13}), (\ref{eq:2-21a})--(\ref{eq:2-21c}),
are completely general, and are applicable to any background
cosmology. One can consider any initial conditions, and the resulting
expressions contain every growing, nongrowing and decaying mode in
general. In practical applications, however, one is interested in the
fastest-growing mode. It is also known that the time dependence of the
fastest-growing mode in higher-order perturbations is approximately
given by $\bm{\varPsi}^{(n)} \propto D^n$. This relation is exact for
the EdS universe, in which $\varOmega_\textrm{m}=1$,
$\varOmega_\Lambda=0$, $D = a$. There are residual time dependencies
in general cosmology, although they are quite small for reasonable
models \cite{Ber94,CLMM95,Mat95,HBCJ95,Tak08,Lee14}. The reason that
the residual time dependencies are small is explained by the structure
of the evolution equation in SPT \cite{MF91,Sco+98}. A similar
argument also applies in the case of LPT as shown below.

Instead of the proper time $t$, we can use the logarithm of linear
growth factor
\begin{equation}
  \tau \equiv \ln D(t),
  \label{eq:2-30}
\end{equation}
as a time variable. In terms of the new variable, the operator of the
type $\hat{\mathcal{T}} - \alpha \pi G\bar{\rho}$, which appears in
the recurrence relations, Eqs.~(\ref{eq:2-3}) and (\ref{eq:2-12}) with
$\alpha=0,4/3,2,4$, is given by
\begin{equation}
  \hat{\mathcal{T}} - \alpha \pi G\bar{\rho} = H^2f^2
  \left[
    \frac{\partial^2}{\partial\tau^2} +
    \left( \frac{3}{2} \frac{\varOmega_\textrm{m}}{f^2} - 1 \right)
    \frac{\partial}{\partial\tau}
    - \frac{3\alpha}{8} \frac{\varOmega_\textrm{m}}{f^2}
  \right],
  \label{eq:2-31}
\end{equation}
where $f = d\ln D/d\ln a$ is the linear growth rate and
$\varOmega_\textrm{m} = 8\pi G\bar{\rho}/3H^2$ is the time dependent
density parameter. The linear growth rate is approximately given by $f
\simeq \varOmega_\textrm{m}^{0.55}$ for flat models \cite{BCHJ95} and
$f \simeq \varOmega_\textrm{m}^{0.6}$ for Friedman models
\cite{Pee80}. If the growth rate is approximated by $f =
\varOmega_\textrm{m}^{1/2}$ in Eq.~(\ref{eq:2-31}), all the
coefficients in the recurrence relations become independent of time;
$n$th-order components of Eqs.~(\ref{eq:2-2}) and (\ref{eq:2-11})
approximately reduce to
\begin{align}
  \left[
    \frac{\partial^2}{\partial\tau^2} + 
    \frac{1}{2} \frac{\partial}{\partial\tau} -
    \frac{3}{2}
  \right]
  \bm{\nabla}\cdot\bm{\varPsi}^{(n)}
  &= - \sum_{m_1 + m_2 = n}
  \varepsilon_{ijk} \varepsilon_{ipq} \varPsi^{(m_1)}_{j,p}
  \left[
    \frac{\partial^2}{\partial\tau^2} + 
    \frac{1}{2} \frac{\partial}{\partial\tau} -
    \frac{3}{4}
  \right] \varPsi^{(m_2)}_{k,q}
  \nonumber\\
  & \quad
  - \frac{1}{2}
  \sum_{m_1+m_2+m_3=n} \varepsilon_{ijk} \varepsilon_{pqr}
  \varPsi^{(m_1)}_{i,p} \varPsi^{(m_2)}_{j,q}
  \left[
    \frac{\partial^2}{\partial\tau^2} + 
    \frac{1}{2} \frac{\partial}{\partial\tau} -
    \frac{1}{2}
  \right] \varPsi^{(m_3)}_{k,r},
  \label{eq:2-32}\\
  \left[
    \frac{\partial^2}{\partial\tau^2} + 
    \frac{1}{2} \frac{\partial}{\partial\tau}
  \right] \bm{\nabla}\times\bm{\varPsi}^{(n)}
  &= \sum_{m_1+m_2=n}
  \bm{\nabla} \varPsi^{(m_1)}_i \times
  \left[
    \frac{\partial^2}{\partial\tau^2} + 
    \frac{1}{2} \frac{\partial}{\partial\tau}
  \right] \bm{\nabla}\varPsi^{(m_2)}_i.
  \label{eq:2-33}
\end{align}
The linear equations with $n=1$ are homogeneous, and the general
solutions are given by a superposition of independent solutions
$e^{\tau} = D$, $e^{-3\tau/2} = D^{-3/2}$ for the longitudinal part,
and $e^{0} = 1$, $e^{-\tau/2} = D^{-1/2}$ for the transverse part. As
a consequence, decaying mode functions in the present approximation
are replaced by $D_-(t) \rightarrow e^{-3\tau/2} = D^{-3/2}$,
$E_-(t) \rightarrow e^{-\tau/2} = D^{-1/2}$. The general solution for
the first-order displacement field is given by Eq.~(\ref{eq:2-19})
with these replacements. In general, the differential equations of
Eqs.~(\ref{eq:2-32}) and (\ref{eq:2-33}) can be recursively solved by
standard methods, e.g., using the Laplace transform, including all the
modes of time-dependence.

When we are interested in the fastest-growing mode, a simple logic of
induction shows that the fastest-growing solutions of the above
equations are given by $\bm{\varPsi}^{(n)} \propto e^{n\tau} = D^n$.
This conclusion is exact in the EdS model, because
$\varOmega_\textrm{m}=f=1$. In general cosmology, even if
approximation $f\simeq\varOmega_\textrm{m}^{1/2}$ is not so accurate
at the present time, it is much more accurate in most of the time
evolution. Thus the ratio $\bm{\varPsi}^{(n)}/D^n$ in exact solutions
is extremely insensitive to background cosmology, even more than what
the approximation $f\simeq \varOmega_\textrm{m}^{1/2}$ would suggest
\cite{BCGS02}.

Substituting the inferred time dependence
$\bm{\varPsi}^{(n)} \propto D^n$ of the fastest-growing mode into
Eqs.~(\ref{eq:2-32}), (\ref{eq:2-33}), the recurrence relations for
$n\geq 2$ with approximate time dependence reduce to
\begin{align}
  \bm{\nabla}\cdot\bm{\varPsi}^{(n)}
  &= -\frac{1}{2} \sum_{\substack{m_1+m_2=n\\m_1\leq m_2}}
  M^{(2)}_{m_1m_2}
  \left[
    1-\frac{4m_1m_2}{(2n+3)(n-1)}
  \right]
  \varepsilon_{ijk} \varepsilon_{ipq} \varPsi^{(m_1)}_{j,p}
  \varPsi^{(m_2)}_{k,q}
\nonumber\\
& \quad
  - \frac{1}{6} \sum_{\substack{m_1+m_2+m_3=n\\m_1\leq m_2\leq m_3}}
  M^{(3)}_{m_1m_2m_3}
  \left[
    1-\frac{4(m_1m_2+m_2m_3+m_3m_1)}{(2n+3)(n-1)}
  \right]
  \varepsilon_{ijk} \varepsilon_{pqr}
  \varPsi^{(m_1)}_{i,p} \varPsi^{(m_2)}_{j,q}
  \varPsi^{(m_3)}_{k,r},
  \label{eq:2-34}\\
  \bm{\nabla}\times\bm{\varPsi}^{(n)}
  &= \sum_{\substack{m_1+m_2=n\\m_1 < m_2}}
  \frac{m_2-m_1}{n}
  \bm{\nabla}\varPsi^{(m_1)}_i \times \bm{\nabla}\varPsi^{(m_2)}_i,
  \label{eq:2-35}
\end{align}
where $M^{(2)}_{m_1m_2}$ and $M^{(3)}_{m_1m_2m_3}$ are multiplicity
factors defined by
\begin{equation}
  M^{(2)}_{m_1m_2} \equiv
  \begin{cases}
      1 & (m_1=m_2) \\
      2 & (m_1 < m_2)
  \end{cases}, \qquad
  M^{(3)}_{m_1m_2m_3} \equiv
  \begin{cases}
      1 & (m_1=m_2=m_3) \\
      3 & (m_1=m_2<m_3) \\
      3 & (m_1<m_2=m_3) \\
      6 & (m_1<m_2<m_3)
  \end{cases}.
  \label{eq:2-35-1}
\end{equation}
The seed values for the above recurrence relation are given by the
fastest-growing mode of Eq.~(\ref{eq:2-19}), i.e., the Zel'dovich
approximation,
\begin{equation}
  \bm{\varPsi}^{(1)} =
  - \bm{\nabla} \mbox{\large{$\triangle$}}^{-1} \delta_\textrm{L}.
  \label{eq:2-35-2}
\end{equation}
Eq.~(\ref{eq:2-34}) can also be derived from Eq.~(\ref{eq:a-11}), an
alternative expression of longitudinal recurrence relations.

The form of Eq.~(\ref{eq:2-35}) is equivalent to the
recurrence relations of Ref.~\cite{Ram12} in the EdS limit which are
derived from the irrotational condition of velocity field in Eulerian
space, $\bm{\nabla}_x\times\bm{v}=\bm{0}$, while our derivation does
{\em not} assume the irrotational condition from the beginning. It is
{\em a consequence} of selecting the fastest-growing mode in LPT that
the Eulerian velocity field is irrotational. In fact, the Eulerian
irrotational condition is equivalent to
$\bm{\nabla}\times\dot{\bm{\varPsi}} =
\bm{\nabla}\varPsi_i\times\bm{\nabla}\dot{\varPsi}_i$,
which is satisfied by Eq.~(\ref{eq:2-35}) with the fastest-growing
mode, $\bm{\varPsi}^{(n)} \propto D^n$. Therefore, unlike most
treatments of LPT in the literature, there is no need for imposing the
irrotational condition $\bm{\nabla}_x\times\bm{v}=\bm{0}$ to begin
with. Solutions with irrotational Eulerian flow form a subclass of
general solutions of LPT, and the fastest-growing mode is in this
subclass. Nevertheless, rotational flows are present in more general
solutions \cite{Buc92}.

The form of Eq.~(\ref{eq:2-34}) is also equivalent to the recurrence
relations of Ref.~\cite{ZF14} in the EdS limit, which are derived from
the Taylor series in the scale factor. Even though the last expansion
scheme is not the same as the one here, they agree with each other in
the special case of the fastest-growing mode in the EdS model. This
agreement does not apply to other cosmological models, because the
linear growth factor is not proportional to the scale factor in
general. In fact, the recurrence relations in the $\Lambda$CDM model
in the last expansion scheme are different from ours \cite{RVF15}.

In Fourier space, the $n$th-order kernel $\tilde{\bm{L}}_n$ defined by
Eq.~(\ref{eq:2-24}) is proportional to $D^n$ in the present
approximation. It is natural to separate this simple time dependence
from the kernels as $\tilde{\bm{L}}_n = D^n\bm{L}_n$, and the newly
defined kernel $\bm{L}_n$ is extremely insensitive to background
cosmology. Eq.~(\ref{eq:2-24}) in this case is represented by
\begin{equation}
  \tilde{\bm{\varPsi}}^{(n)}(\bm{k},t) = \frac{iD^n}{n!}
  \int_{\bm{k}_{1\cdots n}=\bm{k}}
  {\bm{L}}_n(\bm{k}_1,\ldots,\bm{k}_n)
  \delta_0(\bm{k}_1)\cdots\delta_0(\bm{k}_n).
  \label{eq:2-36}
\end{equation}
Defining $S_n = \bm{k}_{1\cdots n}\cdot\bm{L}_n$ and $\bm{T}_n =
-\bm{k}_{1\cdots n}\times\bm{L}_n$, the recurrence relations of
Eqs.~(\ref{eq:2-26a})--(\ref{eq:2-26c}) reduce to
\begin{align}
  S_n(\bm{k}_1,\ldots,\bm{k}_n)
  &= \frac{n!}{2} \sum_{\substack{m_1+m_2=n\\m_1\leq m_2}}
   \frac{M_{m_1m_2}}{m_1! m_2!}
   \left[ 1 - \frac{4m_1m_2}{(2n+3)(n-1)} \right] 
   (\bm{k}_{1\cdots m_1}\times\bm{k}_{(m_1+1)\cdots n})
  \cdot
  \left[
    \bm{L}_{m_1}(\bm{k}_1,\ldots,\bm{k}_{m_1}) \times
    \bm{L}_{m_2}(\bm{k}_{m_1+1},\ldots,\bm{k}_n)
  \right]
\nonumber\\
  & \quad
   - \frac{n!}{6} \sum_{\substack{m_1+m_2+m_3=n\\m_1\leq m_2\leq m_3}}
   \frac{M_{m_1m_2m_3}}{m_1! m_2! m_3!}
  \left[
    1-\frac{4(m_1m_2+m_2m_3+m_3m_1)}{(2n+3)(n-1)}
  \right]
  \left[
    \bm{k}_{1\cdots m_1}\cdot(\bm{k}_{(m_1+1)\cdots(m_1+m_2)}
    \times\bm{k}_{(m_1+m_2+1)\cdots n})
  \right]
\nonumber\\
  & \hspace{11pc}
  \Bigl\{
    \bm{L}_{m_1}(\bm{k}_1,\ldots,\bm{k}_{m_1}) \cdot
    \left[
      \bm{L}_{m_2}(\bm{k}_{m_1+1},\ldots,\bm{k}_{m_1+m_2})\times
      \bm{L}_{m_3}(\bm{k}_{m_1+m_2+1},\ldots,\bm{k}_n)
    \right]
  \Bigr\},
  \label{eq:2-38a}\\
  \bm{T}_n(\bm{k}_1,\ldots,\bm{k}_n)
  & = (n-1)! \sum_{\substack{m_1 + m_2 = n\\m_1 < m_2}}
  \frac{m_2-m_1}{m_1! m_2!}
  (\bm{k}_{1\cdots m_1}\times\bm{k}_{(m_1+1)\cdots n})
  \left[
    \bm{L}_{m_1}(\bm{k}_1,\ldots,\bm{k}_{m_!}) \cdot
    \bm{L}_{m_2}(\bm{k}_{m_1+1},\ldots,\bm{k}_n)
  \right],
  \label{eq:2-38b}\\
  \bm{L}_n(\bm{k}_1,\ldots,\bm{k}_n)
  &=  \frac{1}{{k_{1\cdots n}}^2} 
  \left[
    \bm{k}_{1\cdots n} S_n(\bm{k}_1,\ldots,\bm{k}_n))
    + \bm{k}_{1\cdots n}\times
    \bm{T}_n(\bm{k}_1,\ldots,\bm{k}_n) )
  \right].
  \label{eq:2-38c}
\end{align}
The seed values are given by $\bm{L}_1(\bm{k}) = i\bm{k}/k^2$, i.e.,
$S_1=1$, $\bm{T}_1=\bm{0}$. Practically, the above equations should be
quite useful. The recurrence relations above give unsymmetric kernels,
and symmetric kernels
\begin{equation}
  \bm{L}^\textrm{sym.}_n(\bm{k}_1,\ldots,\bm{k}_n)
  \equiv
  \frac{1}{n!} \sum_{p \in {\cal S}_n}
  \bm{L}_n(\bm{k}_{p(1)},\ldots,\bm{k}_{p(n)}),
  \label{eq:2-39}
\end{equation}
are physically relevant quantities. The vector identities of
Eq.~(\ref{eq:2-29a}) and (\ref{eq:2-29b}) provide an alternative
expression for Eq.~(\ref{eq:2-38a}). For readers' convenience,
recursive solutions up to seventh order are explicitly given in
Appendix~\ref{app:Explicit}.

\section{Conclusions}
\label{sec:Concl}

In this paper, the recurrence relations in the usual framework of LPT
are derived, generalizing the previously known recurrence relations in
LPT. The newly derived relations are self-contained within the usual
framework of LPT, and applicable to any initial condition in a general
cosmological model. The most general recurrence relations are given by
Eqs.~(\ref{eq:2-3}), (\ref{eq:2-12}), (\ref{eq:2-13}) in configuration
space, and Eqs.~(\ref{eq:2-21a})--(\ref{eq:2-21c}) in Fourier space.
The resultant recursive solutions contain not only the fastest-growing
mode but also all the other modes, with arbitrary initial conditions.
In Fourier space, the perturbation kernels of the fastest-growing mode
satisfy the recurrence relations of
Eqs.~(\ref{eq:2-26a})--(\ref{eq:2-26c}).

The above recurrence relations are the most general. With an
approximation for the time dependence, which is very accurate for
realistic models of cosmology, we find simplified recurrence relations
for the fastest-growing mode of LPT. They are given by
Eqs.~(\ref{eq:2-34}), (\ref{eq:2-35}), (\ref{eq:2-13}) in
configuration space, and Eqs.~(\ref{eq:2-38a})--(\ref{eq:2-38c}) in
Fourier space. The time dependence is explicitly solved in the last
cases, and the corresponding recurrence relations are purely
algebraic. Practically, these relations would be the most handy ones
for future applications. Explicit recursive solutions up to seventh
order are given in Appendix~\ref{app:Explicit}.

Unlike most of the previous work, the irrotational condition of
Eulerian velocity field, $\bm{\nabla}_x\times\bm{v} = \bm{0}$, is {\em
  not assumed} throughout this work. However, the recurrence relations
for the transverse part of the fastest-growing mode,
Eqs.~(\ref{eq:2-35}) and (\ref{eq:2-38b}), coincide with those derived
from the irrotational condition. This means that the irrotationality
of Eulerian flow is {\em a consequence} of selecting the
fastest-growing mode in LPT, and there is no need to impose the
condition from the beginning of LPT. Our results for the
fastest-growing mode in the limit of the EdS model are fully
consistent with the previously known recurrence relations.

A straightforward application of this work is to use them for
numerical evaluations of higher-order corrections to statistical
measures of the large-scale structure, such as the power spectrum,
bispectrum, trispectrum, etc. With the machinery of iPT \cite{Mat11}
and/or CLPT \cite{CRW13}, there is a systematic way of calculating
such kinds of statistics from the higher-order LPT, including the
effects of redshift-space distortions, nonlocal bias, and primordial
non-Gaussianity \cite{Mat11,Mat12,YM13,Mat14,YMT14,WRW14}. While two-
or higher-loop nonlinear corrections have been investigated with
Eulerian perturbation theory such as SPT, the applications of LPT have
been limited to one-loop corrections (except Ref.~\cite{OTM11} in
which the SPT/LPT correspondence \cite{Mat11} is used), apparently
because of the lack of recursive solutions in LPT. The recurrence
relations derived in this paper, especially the simplest versions of
Eqs.~(\ref{eq:2-38a})--(\ref{eq:2-38c}), should change the situation
in this respect.

 \begin{acknowledgments}
     I thank Cornelius Rampf for his valuable comments on the
     manuscript. I acknowledge support from the Japanese Ministry of
     Education, Culture, Sports, Science, and Technology, Grant-in-Aid
     for Scientific Research (C) Program, Grants No.~24540267 (2012)
     and No.~15K05074 (2015).
 \end{acknowledgments}

\appendix

\section{An alternative form of recurrence relations for the
    longitudinal part}
\label{app:AltForm}

There is an alternative, but equivalent form of recurrence relations
for the longitudinal part. Using Eq.~(\ref{eq:1-8a}) and the product
rule of Eq.~(\ref{eq:1-9-1b}), one sees Eq.~(\ref{eq:1-10a}) is
equivalent to
\begin{equation}
  \left( \hat{\cal T} - 4\pi G\bar{\rho} \right) (J-1)
  = \varepsilon_{ijk} \varepsilon_{lmn}
  J_{il}\dot{J}_{jm}\dot{J}_{kn}.
  \label{eq:a-4}
\end{equation}
The general solution is formally given by
\begin{equation}
  J - 1 =
  A_+ D_+ + A_- D_-
  + \varepsilon_{ijk} \varepsilon_{lmn}
  \left(\hat{\cal T} - 4\pi G\bar{\rho}\right)^{-1}
  J_{il}\dot{J}_{jm}\dot{J}_{kn},
  \label{eq:a-9}
\end{equation}
where $A_\pm(\bm{q})$ are integration constants of time. Substituting
Eqs.~(\ref{eq:1-6}), (\ref{eq:1-8a}) into this equation, we have
\begin{equation}
  \bm{\nabla}\cdot\bm{\varPsi}
  = A_+ D_+ + A_- D_-
  - \frac{1}{2}
  \varepsilon_{ijk}\varepsilon_{ilm}
  \varPsi_{j,l} \varPsi_{k,m}
  - \frac{1}{6} \varepsilon_{ijk}\varepsilon_{lmn}
  \varPsi_{i,l} \varPsi_{j,m} \varPsi_{k,n}
  + \left( \hat{\cal T} - 4\pi G\bar{\rho} \right)^{-1}
  \left[
    \varepsilon_{ijk} \varepsilon_{ilm}
    \dot{\varPsi}_{jl}\dot{\varPsi}_{km}
    + \varepsilon_{ijk} \varepsilon_{lmn}
    \varPsi_{il}\dot{\varPsi}_{jm}\dot{\varPsi}_{kn}
  \right].
  \label{eq:a-10}
\end{equation}
The linear solution is given by
\begin{equation}
  \bm{\nabla}\cdot\bm{\varPsi}^{(1)} = A_+ D_+ + A_- D_-,
  \label{eq:a-11}
\end{equation}
and extracting the particular order $n \geq 2$, we have
\begin{multline}
  \bm{\nabla}\cdot\bm{\varPsi}^{(n)}
  = - \frac{1}{2} \sum_{m_1+m_2=n} \varepsilon_{ijk}\varepsilon_{ilm}
  \varPsi^{(m_1)}_{j,l} \varPsi^{(m_2)}_{k,m}
  - \frac{1}{6} \sum_{m_1+m_2+m_3=n} \varepsilon_{ijk}\varepsilon_{lmn}
  \varPsi^{(m_1)}_{i,l} \varPsi^{(m_2)}_{j,m} \varPsi^{(m_3)}_{k,n}
\\
  + \left( \hat{\cal T} - 4\pi G\bar{\rho} \right)^{-1}
  \left[
    \sum_{m_1+m_2=n}
    \varepsilon_{ijk} \varepsilon_{ilm}
    \dot{\varPsi}^{(m_1)}_{j,l}\dot{\varPsi}^{(m_2)}_{k,m}
    + \sum_{m_1+m_2+m_3=n}
    \varepsilon_{ijk} \varepsilon_{lmn} \varPsi^{(m_1)}_{i,l}
    \dot{\varPsi}^{(m_2)}_{j,m} \dot{\varPsi}^{(m_3)}_{k,n}
  \right].
  \label{eq:a-12}
\end{multline}
The above equation is equivalent to Eq.~(\ref{eq:2-3}), as
straightforwardly confirmed by using the product rules of
Eqs.~(\ref{eq:1-9-1a}) and (\ref{eq:1-9-1b}).

\section{Explicit solutions up to seventh order}
\label{app:Explicit}


Explicitly writing down the perturbation kernel of each order is
straightforward by applying the recurrence relations derived in the
main text. For readers' convenience, we manifestly show the explicit
kernel functions of the fastest-growing mode with approximate
time dependence, up to seventh order in this Appendix. The
seventh-order perturbation theory is required in calculating, e.g.,
three-loop corrections to the power spectrum, etc. As in the main
text, we apply notations
\begin{align}
  S_n(\bm{k}_1,\ldots,\bm{k}_n) &\equiv
  \bm{k}_{1\cdots n}\cdot \bm{L}_n(\bm{k}_1,\ldots,\bm{k}_n),
  \label{eq:b-2a}\\
  \bm{T}_n(\bm{k}_1,\ldots,\bm{k}_n) &\equiv
  - \bm{k}_{1\cdots n}\times \bm{L}_n(\bm{k}_1,\ldots,\bm{k}_n),
  \label{eq:b-2b}
\end{align}
and results of these functions are presented below. The total kernel
is given by
\begin{equation}
  \bm{L}_n(\bm{k}_1,\ldots,\bm{k}_n) = 
  \frac{1}{{{k}_{1\cdots n}}^2}
  \left[
    \bm{k}_{1\cdots n} S_n(\bm{k}_1,\ldots,\bm{k}_n) + 
    \bm{k}_{1\cdots n} \times \bm{T}_n(\bm{k}_1,\ldots,\bm{k}_n)
  \right].
  \label{eq:b-3}
\end{equation}
from the above functions. For simplicity, the following results are
unsymmetric with respect to their arguments. The symmetrization
procedure, Eq.~(\ref{eq:2-39}) should be applied to obtain symmetric
kernels.

To present the results, it is convenient to define the following
functions,
\begin{align}
  U(\bm{k}_1,\bm{k}_2)
  &= \frac{\left|\bm{k}_1\times\bm{k}_2\right|^2}{{k_1}^2{k_2}^2}
    = 1 - \left(\frac{\bm{k}_1\cdot\bm{k}_2}{k_1 k_2}\right)^2,
  \label{eq:b-4a}\\
  V(\bm{k}_1,\bm{k}_2,\bm{k}_3)
  &= \frac{\left|\bm{k}_1\cdot(\bm{k}_2\times\bm{k}_3)\right|^2}
    {{k_1}^2{k_2}^2{k_2}^3}
  = 1 - \left(\frac{\bm{k}_1\cdot\bm{k}_2}{k_1 k_2}\right)^2
  - \left(\frac{\bm{k}_2\cdot\bm{k}_3}{k_2 k_3}\right)^2
  - \left(\frac{\bm{k}_3\cdot\bm{k}_1}{k_3 k_1}\right)^2
  + 2
  \frac{(\bm{k}_1\cdot\bm{k}_2)(\bm{k}_2\cdot\bm{k}_3)(\bm{k}_3\cdot\bm{k}_1)}
  {{k_1}^2{k_2}^2{k_3}^2},
  \label{eq:b-4b}\\
  \bm{W}(\bm{k}_1,\bm{k}_2)
  &= \frac{(\bm{k}_1\times\bm{k}_2)(\bm{k}_1\cdot\bm{k}_2)}{{k_1}^2{k_2}^2}.
  \label{eq:b-4c}
\end{align}

The recurrence relations enable us to deduce perturbation kernels of
arbitrary order, once the first-order solution is given. As we are
interested in the fastest-growing solution, the first-order solution
is given by the Zel'dovich approximation, $\bm{L}_1(\bm{k}) =
\bm{k}/k^2$, i.e.,
\begin{equation}
  S_1(\bm{k}) = 1, \quad
  \bm{T}_1(\bm{k}) = \bm{0}.
  \label{eq:b-4}
\end{equation}
The higher-order solutions are recursively derived from
Eqs.~(\ref{eq:2-38a}) and (\ref{eq:2-38b}). The second-order solution
is given by
\begin{equation}
  S_2(\bm{k}_1,\bm{k}_2) = \frac{3}{7} U(\bm{k}_1,\bm{k}_2), \quad
  \bm{T}_2(\bm{k}_1,\bm{k}_2) = \bm{0}.
  \label{eq:b-5}
\end{equation}
The third-order solution is given by
\begin{align}
  S_3(\bm{k}_1,\bm{k}_2,\bm{k}_3)
  &= \frac{5}{3} U(\bm{k}_1,\bm{k}_{23})S_2(\bm{k}_2,\bm{k}_3)
    - \frac{1}{3} V(\bm{k}_1,\bm{k}_2,\bm{k}_3),
  \label{eq:b-8a}\\
  \bm{T}_3(\bm{k}_1,\bm{k}_2,\bm{k}_3)
  &= \bm{W}(\bm{k}_1,\bm{k}_{23}) S_2(\bm{k}_2,\bm{k}_3).
  \label{eq:b-8b}
\end{align}
Transverse parts $\bm{T}_n$ appear only from the third order
($n\geq 3$) in the fastest-growing-mode solutions. This property is a
consequence of the fastest-growing mode with approximate
time dependence, and does not apply in general solutions \cite{BE93},
as seen from Eqs.~(\ref{eq:2-11-4}) and (\ref{eq:2-12}). The
fourth-order solution is given by
\begin{align}
  S_4(\bm{k}_1,\ldots,\bm{k}_4)
  &= \frac{28}{11}
  \Biggl[
    U(\bm{k}_1,\bm{k}_{234})S_3(\bm{k}_2,\bm{k}_3,\bm{k}_4)
    - \bm{W}(\bm{k}_1,\bm{k}_{234})\cdot\bm{T}_3(\bm{k}_2,\bm{k}_3,\bm{k}_4)
  \Biggr]
  \nonumber\\
  & \quad
  + \frac{17}{11}
  U(\bm{k}_{12},\bm{k}_{34}) S_2(\bm{k}_1,\bm{k}_2) S_2(\bm{k}_3,\bm{k}_4)
  - \frac{26}{11} V(\bm{k}_1,\bm{k}_2,\bm{k}_{34}) S_2(\bm{k}_3,\bm{k}_4),
  \label{eq:b-10a}\\
  \bm{T}_4(\bm{k}_1,\ldots,\bm{k}_4)
  &= 2
  \Biggl[
    \bm{W}(\bm{k}_1,\bm{k}_{234}) S_3(\bm{k}_2,\bm{k}_3,\bm{k}_4)
    + \frac{\bm{k}_1\times\bm{k}_{234}}{{k_1}^2{k_{234}}^2}
    (\bm{k}_1\times\bm{k}_{234})\cdot\bm{T}_3(\bm{k}_2,\bm{k}_3,\bm{k}_4)
    \Biggr].
  \label{eq:b-10b}
\end{align}
Substituting Eqs.~(\ref{eq:b-4})--(\ref{eq:b-8b}) into the above
equation, explicit solutions in the EdS universe derived in
Ref.~\cite{RB12} are exactly reproduced.

Solutions of LPT for fifth or even higher orders are not found in the
literature. The fifth-order solution from the recurrence relations is
given by
\begin{align}
  S_5(\bm{k}_1,\ldots,\bm{k}_5)
  &= \frac{45}{13}
  \Biggl[
    U(\bm{k}_1,\bm{k}_{2345})S_4(\bm{k}_2,\ldots,\bm{k}_5)
    - \bm{W}(\bm{k}_1,\bm{k}_{2345})\cdot
    \bm{T}_4(\bm{k}_2,\ldots,\bm{k}_5)
  \Biggr]
  \nonumber\\
  & \quad
  + \frac{70}{13} S_2(\bm{k}_1,\bm{k}_2)
  \Biggl[
    U(\bm{k}_{12},\bm{k}_{345}) S_3(\bm{k}_3,\bm{k}_4,\bm{k}_5)
    - \bm{W}(\bm{k}_{12},\bm{k}_{345})\cdot
    \bm{T}_3(\bm{k}_3,\bm{k}_4,\bm{k}_5)
  \Biggr]
  \nonumber\\
  & \quad
  - \frac{60}{13}
  \left\{
    V(\bm{k}_1,\bm{k}_2,\bm{k}_{345})
    S_3(\bm{k}_3,\bm{k}_4,\bm{k}_5)
    + \frac{(\bm{k}_1\times\bm{k}_2)\cdot\bm{k}_{345}}
    {{k_1}^2{k_2}^2{k_{345}}^2}
    \left[
      (\bm{k}_1\times \bm{k}_2)\times\bm{k}_{345}
    \right]\cdot\bm{T}_3(\bm{k}_3,\bm{k}_4,\bm{k}_5)
  \right\}
  \nonumber\\
  & \quad
  - \frac{75}{13} V(\bm{k}_1,\bm{k}_{23},\bm{k}_{45})
  S_2(\bm{k}_2,\bm{k}_3) S_2(\bm{k}_4,\bm{k}_5),
  \label{eq:b-12a}\\
  \bm{T}_5(\bm{k}_1,\ldots,\bm{k}_5)
  &= 3
  \left[
      \bm{W}(\bm{k}_1,\bm{k}_{2345}) S_4(\bm{k}_2,\ldots,\bm{k}_5)
    + \frac{\bm{k}_1\times\bm{k}_{2345}}{{k_1}^2{k_{2345}}^2}
    (\bm{k}_1\times\bm{k}_{2345})\cdot\bm{T}_4(\bm{k}_2,\ldots,\bm{k}_5)
  \right]
  \nonumber\\
  & \quad
  + 2 S_2(\bm{k}_1,\bm{k}_2)
  \left[
      \bm{W}(\bm{k}_{12},\bm{k}_{345}) S_3(\bm{k}_3,\bm{k}_4,\bm{k}_5)
    + \frac{\bm{k}_{12}\times\bm{k}_{345}}{{k_{12}}^2{k_{345}}^2}
    (\bm{k}_{12}\times\bm{k}_{345})\cdot\bm{T}_3(\bm{k}_3,\bm{k}_4,\bm{k}_5)
  \right].
  \label{eq:b-12b}
\end{align}
The sixth-order solution is given by
\begin{align}
  S_6(\bm{k}_1,\ldots,\bm{k}_6)
  &= \frac{22}{5}
  \Biggl[
    U(\bm{k}_1,\bm{k}_{23456})S_5(\bm{k}_2,\ldots,\bm{k}_6)
    - \bm{W}(\bm{k}_1,\bm{k}_{23456})\cdot
    \bm{T}_5(\bm{k}_2,\ldots,\bm{k}_6)
  \Biggr]
  \nonumber\\
  & \quad
  + \frac{43}{5} S_2(\bm{k}_1,\bm{k}_2)
  \Biggl[
    U(\bm{k}_{12},\bm{k}_{3456}) S_4(\bm{k}_3,\ldots,\bm{k}_6)
    - \bm{W}(\bm{k}_{12},\bm{k}_{3456})\cdot
    \bm{T}_4(\bm{k}_3,\ldots,\bm{k}_6)
  \Biggr]
  \nonumber\\
  & \quad
  + \frac{26}{5}
  \Biggl[
  S_3(\bm{k}_1,\bm{k}_2,\bm{k}_3)
  \left[
    U(\bm{k}_{123},\bm{k}_{456})
    S_3(\bm{k}_4,\bm{k}_5,\bm{k}_6)
    - 2 \bm{W}(\bm{k}_{123},\bm{k}_{456})
      \cdot\bm{T}_3(\bm{k}_4,\bm{k}_5,\bm{k}_6)
    \right]
  \nonumber\\
  & \qquad\qquad
    + \frac{\bm{k}_{123}\times\bm{k}_{456}}{{k_{123}}^2{k_{456}}^2}\cdot
    \Bigl\{
      \left[
        \bm{k}_{123}\times\bm{T}_3(\bm{k}_1,\bm{k}_2,\bm{k}_3)
      \right]
      \times
      \left[
        \bm{k}_{456}\times\bm{T}_3(\bm{k}_4,\bm{k}_5,\bm{k}_6)
      \right]
    \Bigr\}
  \Biggr]
  \nonumber\\
  & \quad
  - \frac{39}{5}
  \left\{
    V(\bm{k}_1,\bm{k}_2,\bm{k}_{3456}) S_4(\bm{k}_3,\ldots,\bm{k}_6)
    + \frac{(\bm{k}_1\times\bm{k}_2)\cdot\bm{k}_{3456}}
    {{k_1}^2{k_2}^2{k_{3456}}^2}
    \left[
      (\bm{k}_1\times \bm{k}_2)\times\bm{k}_{3456}
    \right]
    \cdot\bm{T}_4(\bm{k}_3,\ldots,\bm{k}_6)
  \right\}
  \nonumber\\
  & \quad
  - \frac{124}{5} S_2(\bm{k}_2,\bm{k}_3)
  \left\{
    V(\bm{k}_1,\bm{k}_{23},\bm{k}_{456})
    S_3(\bm{k}_4,\bm{k}_5,\bm{k}_6)
    + \frac{(\bm{k}_1\times\bm{k}_{23})\cdot\bm{k}_{456}}
    {{k_1}^2{k_{23}}^2{k_{456}}^2}
    \left[
      (\bm{k}_1\times \bm{k}_{23})\times\bm{k}_{456}
    \right]
    \cdot\bm{T}_3(\bm{k}_4,\bm{k}_5,\bm{k}_6)]
  \right\}
  \nonumber\\
  & \quad
  - \frac{27}{5} V(\bm{k}_{12},\bm{k}_{34},\bm{k}_{56})
  S_2(\bm{k}_1,\bm{k}_2) S_2(\bm{k}_3,\bm{k}_4) S_2(\bm{k}_5,\bm{k}_6),
  \label{eq:b-13a}\\
  \bm{T}_6(\bm{k}_1,\ldots,\bm{k}_6)
  &= 4
  \Biggl[
      \bm{W}(\bm{k}_1,\bm{k}_{23456}) S_5(\bm{k}_2,\ldots,\bm{k}_6)
    + \frac{\bm{k}_1\times\bm{k}_{23456}}{{k_1}^2{k_{23456}}^2}
      (\bm{k}_1\times\bm{k}_{23456})\cdot\bm{T}_5(\bm{k}_2,\ldots,\bm{k}_6)
  \Biggr]
  \nonumber\\
  & \quad
  + 5 S_2(\bm{k}_1,\bm{k}_2)
  \Biggl[
      \bm{W}(\bm{k}_{12},\bm{k}_{3456}) S_4(\bm{k}_3,\ldots,\bm{k}_6)
    + \frac{\bm{k}_{12}\times\bm{k}_{3456}}{{k_{12}}^2{k_{3456}}^2}
    (\bm{k}_{12}\times\bm{k}_{3456})
    \cdot\bm{T}_4(\bm{k}_3,\ldots,\bm{k}_6)
  \Biggr].
  \label{eq:b-13b}
\end{align}
The seventh-order solution is given by
\begin{align}
  S_7(\bm{k}_1,\ldots,\bm{k}_7)
  &= \frac{91}{17}
  \Biggl[
    U(\bm{k}_1,\bm{k}_{234567})S_6(\bm{k}_2,\ldots,\bm{k}_7)
    - \bm{W}(\bm{k}_1,\bm{k}_{234567})\cdot
    \bm{T}_6(\bm{k}_2,\ldots,\bm{k}_7)
  \Biggr]
  \nonumber\\
  & \quad
  + \frac{217}{17} S_2(\bm{k}_1,\bm{k}_2)
  \Biggl[
    U(\bm{k}_{12},\bm{k}_{34567}) S_5(\bm{k}_3,\ldots,\bm{k}_7)
    - \bm{W}(\bm{k}_{12},\bm{k}_{34567})\cdot
    \bm{T}_5(\bm{k}_3,\ldots,\bm{k}_7)
  \Biggr]
  \nonumber\\
  & \quad
  + \frac{315}{17}
  \Biggl[
    U(\bm{k}_{123},\bm{k}_{4567})
    S_3(\bm{k}_1,\bm{k}_2,\bm{k}_3)
    S_4(\bm{k}_4,\ldots,\bm{k}_7)
  \nonumber\\
  & \qquad\qquad
    - \bm{W}(\bm{k}_{123},\bm{k}_{4567})\cdot
    \left[
      S_3(\bm{k}_1,\bm{k}_2,\bm{k}_3)
      \bm{T}_4(\bm{k}_4,\ldots,\bm{k}_7)
      - \bm{T}_3(\bm{k}_1,\bm{k}_2,\bm{k}_3)
      S_4(\bm{k}_5,\ldots,\bm{k}_7)
    \right]
  \nonumber\\
  & \qquad\qquad
    + \frac{\bm{k}_{123}\times\bm{k}_{4567}}{{k_{123}}^2{k_{4567}}^2}\cdot
    \Bigl\{
      \left[
        \bm{k}_{123}\times\bm{T}_3(\bm{k}_1,\bm{k}_2,\bm{k}_3)
      \right]
      \times
      \left[
        \bm{k}_{4567}\times\bm{T}_4(\bm{k}_4,\ldots,\bm{k}_7)
      \right]
    \Bigr\}
  \Biggr]
  \nonumber\\
  & \quad
  - \frac{203}{17}
  \left\{
    V(\bm{k}_1,\bm{k}_2,\bm{k}_{34567}) S_4(\bm{k}_3,\ldots,\bm{k}_7)
    + \frac{(\bm{k}_1\times\bm{k}_2)\cdot\bm{k}_{34567}}
    {{k_1}^2{k_2}^2{k_{34567}}^2}
    \left[
      (\bm{k}_1\times \bm{k}_2)\times\bm{k}_{34567}
    \right]
    \cdot\bm{T}_5(\bm{k}_3,\ldots,\bm{k}_7)
  \right\}
  \nonumber\\
  & \quad
  - \frac{805}{17} S_2(\bm{k}_2,\bm{k}_3)
  \left\{
    V(\bm{k}_1,\bm{k}_{23},\bm{k}_{4567})
    S_4(\bm{k}_4,\ldots,\bm{k}_7)
    + \frac{(\bm{k}_1\times\bm{k}_{23})\cdot\bm{k}_{4567}}
    {{k_1}^2{k_{23}}^2{k_{4567}}^2}
    \left[
      (\bm{k}_1\times \bm{k}_{23})\times\bm{k}_{4567}
    \right]
    \cdot\bm{T}_4(\bm{k}_4,\ldots,\bm{k}_7)]
  \right\}
  \nonumber\\
  & \quad
  - \frac{490}{17}
  \Biggl[
    V(\bm{k}_1,\bm{k}_{234},\bm{k}_{567})
    S_3(\bm{k}_2,\bm{k}_3,\bm{k}_4)
    S_3(\bm{k}_5,\bm{k}_6,\bm{k}_7)
  \nonumber\\
  & \qquad\qquad
    + 2S_3(\bm{k}_2,\bm{k}_3,\bm{k}_4)
    \frac{(\bm{k}_1\times\bm{k}_{234})\cdot\bm{k}_{567}}
    {{k_1}^2{k_{234}}^2{k_{567}}^2}
    \left[
      (\bm{k}_1\times \bm{k}_{234})\times\bm{k}_{567}
    \right]
    \cdot\bm{T}_3(\bm{k}_5,\bm{k}_6,\bm{k}_7)]
  \nonumber\\
  & \qquad\qquad
    + \frac{(\bm{k}_1\times\bm{k}_{234})\cdot\bm{k}_{567}}
    {{k_1}^2{k_{234}}^2{k_{567}}^2}\bm{k}_1\cdot
    \Bigl\{
      \left[
        \bm{k}_{234}\times\bm{T}_3(\bm{k}_2,\bm{k}_3,\bm{k}_4)
      \right]
      \times
      \left[
        \bm{k}_{567}\times\bm{T}_3(\bm{k}_5,\bm{k}_6,\bm{k}_7)
      \right]
    \Bigr\}
  \Biggr]
  \nonumber\\
  & \quad
  - \frac{665}{17} S_2(\bm{k}_1,\bm{k}_2) S_2(\bm{k}_3,\bm{k}_4)
  \Biggl\{
    V(\bm{k}_{12},\bm{k}_{34},\bm{k}_{567})
    S_3(\bm{k}_5,\bm{k}_6,\bm{k}_7)
  \nonumber\\
  & \hspace{12pc}
    + \frac{(\bm{k}_{12}\times\bm{k}_{34})\cdot\bm{k}_{567}}
    {{k_{12}}^2{k_{34}}^2{k_{567}}^2}
    \left[
      (\bm{k}_{12}\times \bm{k}_{34})\times\bm{k}_{567}
    \right]
    \cdot\bm{T}_3(\bm{k}_5,\bm{k}_6,\bm{k}_7)]
  \Biggr\},
  \label{eq:b-14a}\\
  \bm{T}_7(\bm{k}_1,\ldots,\bm{k}_7)
  &= 5
  \left[
      \bm{W}(\bm{k}_1,\bm{k}_{234567}) S_6(\bm{k}_2,\ldots,\bm{k}_7)
    + \frac{\bm{k}_1\times\bm{k}_{234567}}{{k_1}^2{k_{234567}}^2}
    (\bm{k}_1\times\bm{k}_{234567})\cdot\bm{T}_6(\bm{k}_2,\ldots,\bm{k}_7)
  \right]
  \nonumber\\
  & \quad
  + 9
  S_2(\bm{k}_1,\bm{k}_2)
  \Bigl[
      \bm{W}(\bm{k}_{12},\bm{k}_{34567}) S_5(\bm{k}_3,\ldots,\bm{k}_7)
    + \frac{\bm{k}_{12}\times\bm{k}_{34567}}{{k_{12}}^2{k_{34567}}^2}
    (\bm{k}_{12}\times\bm{k}_{34567})
    \cdot\bm{T}_5(\bm{k}_3,\ldots,\bm{k}_7)
  \Bigr].
  \nonumber\\
  & \quad
  + 5
  \Biggl[
      \bm{W}(\bm{k}_{123},\bm{k}_{4567})
      S_3(\bm{k}_1,\bm{k}_2,\bm{k}_3) S_4(\bm{k}_4,\ldots,\bm{k}_7)
  \nonumber\\
  & \hspace{2.5pc}
    + \frac{\bm{k}_{123}\times\bm{k}_{4567}}{{k_{123}}^2{k_{4567}}^2}
    \Bigl\{
      (\bm{k}_{123}\times\bm{k}_{4567})\cdot
      \left[
          S_3(\bm{k}_1,\bm{k}_2,\bm{k}_3) 
          \bm{T}_4(\bm{k}_4,\ldots,\bm{k}_7)
          - \bm{T}_3(\bm{k}_1,\bm{k}_2,\bm{k}_3)
          S_4(\bm{k}_4,\ldots,\bm{k}_7) 
      \right]
      \nonumber\\
      & \hspace{8pc}
      +
      \left[
          \bm{k}_{123}\times\bm{T}_3(\bm{k}_1,\bm{k}_2,\bm{k}_3)
      \right]\cdot
      \left[
          \bm{k}_{4567}\times\bm{T}_4(\bm{k}_4,\ldots,\bm{k}_7)
      \right]
    \Bigr\}
  \Biggr].
  \label{eq:b-14b}
\end{align}
Continuing this kind of calculation and writing down similar
expressions for $n\geq 8$ is straightforward and not difficult thanks
to the recurrence relations.



\renewcommand{\apj}{Astrophys.~J. }
\newcommand{\aap}{Astron.~Astrophys. }
\newcommand{\aj}{Astron.~J. }
\newcommand{\apjl}{Astrophys.~J.~Lett. }
\newcommand{\apjs}{Astrophys.~J.~Suppl.~Ser. }
\newcommand{\apss}{Astrophys.~Space Sci. }
\newcommand{\jcap}{J.~Cosmol.~Astropart.~Phys. }
\newcommand{\jhep}{JHEP }
\newcommand{\mnras}{Mon.~Not.~R.~Astron.~Soc. }
\newcommand{\mpla}{Mod.~Phys.~Lett.~A }
\newcommand{\pasj}{Publ.~Astron.~Soc.~Japan }
\newcommand{\physrep}{Phys.~Rep. }
\newcommand{\ptp}{Progr.~Theor.~Phys. }
\newcommand{\ptep}{Prog.~Theor.~Exp.~Phys. }
\newcommand{\jetp}{JETP }


\end{document}